**DECISION SCIENCES INSTITUTE**
Understanding What Drives Bitcoin Trading Activities

Natalia Jerdack
Farmer School of Business, Miami University
Email: jerdacn2@miamioh.edu

Akmaral Dauletbek
Farmer School of Business, Miami University
Email: dauleta@miamioh.edu

Meredith Divine
Farmer School of Business, Miami University
Email: divinemn@miamioh.edu

Michael Hult
Farmer School of Business, Miami University
Email: hultm@miamioh.edu

Arthur Carvalho
Farmer School of Business, Miami University
Email: arthur.carvalho@miamioh.edu

**ABSTRACT**

Cryptocurrencies have gained tremendous popularity over the past few years. The purpose of this study is to try to understand the factors that are driving cryptocurrency-related trading activities. Focusing on the well-established cryptocurrency called Bitcoin, we find that online search popularity and the volume of trade in unrelated stock markets positively and negatively, respectively, influence Bitcoin trading volume. We also find no statistical evidence that the underlying sentiment behind relevant financial news influence Bitcoin trading volume. We believe these results might be of great value to investors interested in cryptocurrencies and might instigate further research on this topic.

**INTRODUCTION**

The year of 2009 saw the birth of a revolutionary concept, namely the online, fully decentralized currency called *Bitcoin* (Nakamoto, 2008). As a payment system, the transactions involving bitcoins are recorded in a public, distributed ledger that requires no intermediaries such as a central bank. That distributed ledger, called *Blockchain*, is heavily dependent on concepts and ideas from the cryptography field, which makes it a member of a new family of information technologies called cryptotechnologies. Due to a similar reason, Bitcoin is now considered a member of a family of currencies called *cryptocurrencies*.

Since its first release, Bitcoin has gained tremendous popularity over the years and exploded in its valuation. For example, Figure 1 shows the value of one Bitcoin in US dollar from July 18[th], 2010 to May 15[th], 2018. One can immediately see that there was a huge spike in prices in 2017. In particular, Bitcoin price peaked at $ 19,343.04 on December 16[th], 2017.

Figure 1. Bitcoin Prices in US Dollar from July 18th, 2010 to May 15th, 2018.

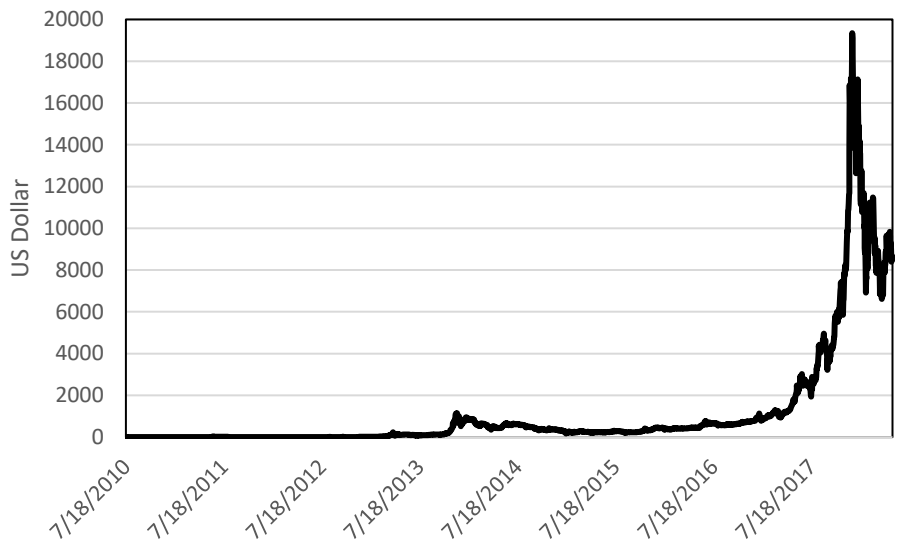

Following the success of Bitcoin, several other cryptocurrencies raised money through initial coin offerings (also known as ICOs) and are now publicly available for trading. As of May 18[th], 2018, the website *coinmarketcap.com* listed a total of 1,593 cryptocurrencies having a combined market cap of $369,691,771,684. In all fairness, after the current hype around cryptocurrencies dies down, it is unlikely that all those cryptocurrencies will stand the test of time. The question that arises is then: which cryptocurrencies will survive? Answering this question is crucial for low-risk-tolerance investors and/or investors considering long-term cryptocurrency investment strategies.

One way of determining whether a certain cryptocurrency will stand the test of time is by looking at trading volume. In particular, one can take inexistent or very low trading activities as a proxy for the lack of interest in the underlying cryptocurrency. In this paper, we try to understand some of the factors that might influence trading activities associated with cryptocurrencies. Specifically, we focus on the potential factors that drive Bitcoin trading volume due to the same being currently the most well-established cryptocurrency.

Since cryptocurrencies are online coins, it might be just natural that trading volume is partially driven by the online popularity of a cryptocurrency. Our proxy for online popularity is the frequency with which online searches include the name of a cryptocurrency. As we elaborate on later, we use data from Google Trends to measure search frequency. That said, our first hypothesis is:

Hypothesis #1: *online search frequency positively correlates with Bitcoin trading volume.*

We next hypothesize that other trading activities might influence the volume of cryptocurrency-related trade. For example, one can argue that when a certain market (*e.g.*, a stock market) is attractive, then less resources might be allocated to other trading activities. To test this idea, we measure how the trading volume associated with the stock market index known as the Dow Jones Industrial Average (DJIA) influences Bitcoin trading volume. Our second hypothesis is then:

Hypothesis #2: *the trading volume in non-cryptocurrency financial markets negatively correlates with Bitcoin trading volume.*

Our last hypothesis relates to the influence of financial news on the trading volume concerning cryptocurrencies. It comes as no surprise that financial news heavily influence investors (Barber & Odean, 2007; Fang & Peress, 2009; Engelberg & Parsons, 2011). Recent financial news have mixed feelings when it comes to cryptocurrencies. On the one hand, there are positive news around the acceptance of cryptocurrencies and their valuation gains. On the other hand, there are also several reports on how the anonymity aspect of some cryptocurrencies are making them very suitable to be used for the payments of illegal actives. Since the sentiment behind the underlying news is mixed and, generally speaking, cryptocurrencies are growing in value, we then hypothesize that financial news have no influence on the trading volume of cryptocurrencies. To test this hypothesis, we analyze how the sentiment behind the news published on the Facebook page called *Bitcoin Chart* affects Bitcoin trading volume. Our formal hypothesis is then:

Hypothesis #3: *the sentiment behind cryptocurrency-related news does not significantly affect Bitcoin trading volume*.

In the following section, we explain how we collect the data relevant to the testing of the above hypotheses. This is followed by an explanation of how we analyze the collected data. We finally conclude by elaborating on the implications of the obtained results and how they relate to the relevant literature.

## DATA COLLECTION AND PREPARATION

The central variable in our study, henceforth called *Bitcoin_Volume*, measures Bitcoin trading volume. We collected its values from the website *blockchain.info* (Blockchain, 2018). The collected data covers the period of time between July 24th, 2017 and April 19th, 2018, which captures the moment in time when Bitcoin exploded in valuation (see Figure 1 and 2). The resulting 270 observations correspond to the number of daily confirmed Bitcoin transactions. To test Hypothesis 1, 2, and 3, we also collected data from Google Trends, DJIA, and Facebook, as we explain next.

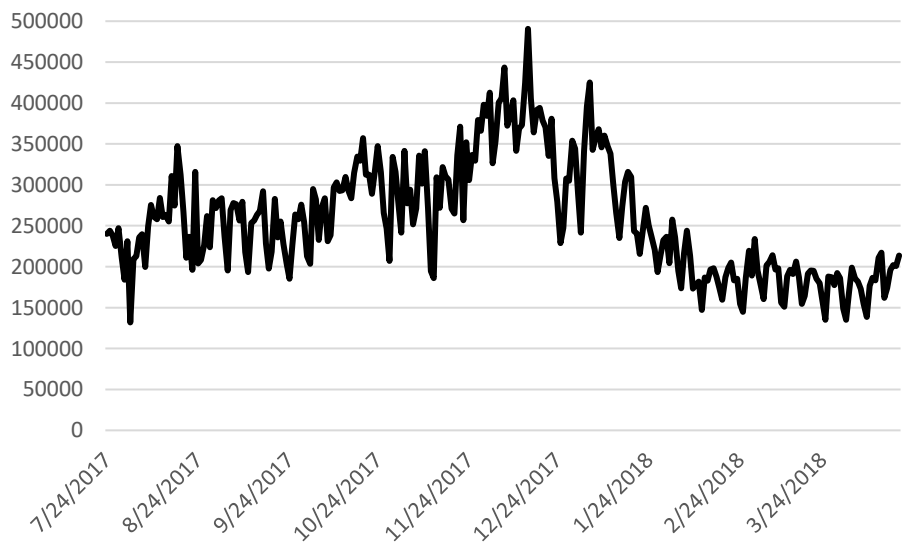

Figure 2. Number of Daily Confirmed Bitcoin Transactions.

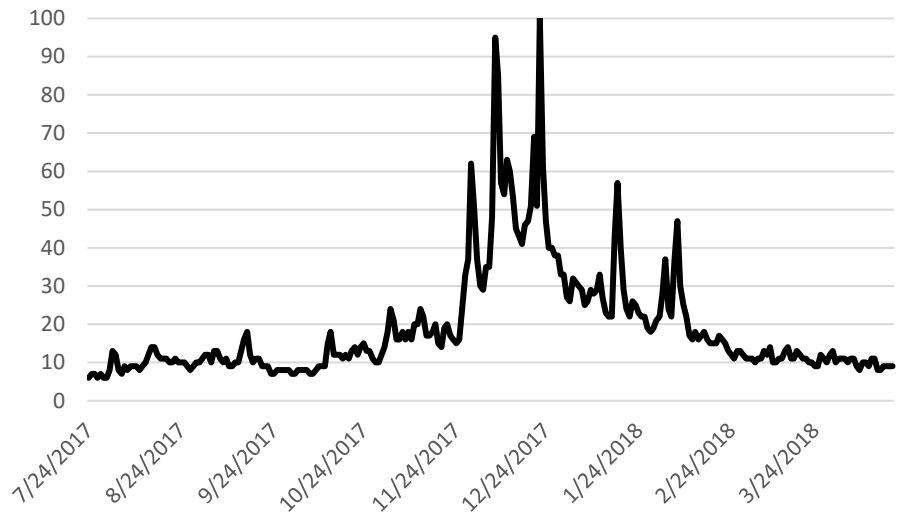

Figure 3. Google Trends Regarding the Term 'Bitcoin'.

### Google Trends

The next source of data is Google Trends (*trends.google.com*). The purpose of the collected variable, henceforth called *Gtrend*, is to determine the online popularity of the term "Bitcoin" over time. Specifically, Google Trends determines the "interest over time" for a specific search term by dividing the number of searches of that term by the total number of all searches done on Google at a given point in time. The resulting numbers are then scaled on a range of 0 to 100 based on the term's proportion to all searches on all topics. In our work, we consider daily searches done by users in the United States of America. That said, we obtained 270 values between 0 and 100 that correspond to how popular the term 'Bitcoin' was during the period of time between July 24th, 2017 and April 19th, 2018. Figure 3 illustrates the obtained data.

### Dow Jones Industrial Average

The next variable we collected, henceforth called *DJIA_Volume*, is about values representing daily trading volumes associated with the Dow Jones Industrial Average (DJIA) index. This data set was collected from Yahoo Finance (Yahoo, 2018). DJIA determines how 30 major American companies have traded on the NASDAQ and NYSE stock markets. Such an index includes very diverse companies, *e.g.*, Apple, Boeing, Caterpillar, Goldman Sachs, IBM, Nike, Walmart, among others. Since the underlying stock markets are officially closed on the weekends, we were only able to collect 187 values during the period of time between July 24th, 2017 and April 19th, 2018.

### Facebook

We finally collected financial news related to Bitcoin published on the Facebook public page called *Bitcoin Chart* (Facebook 2018). At the time of writing, that page has the highest number of followers among open Bitcoin pages on Facebook with a total of 402,562 followers. In total, we collected 694 Bitcoin-related snippets across 229 different days between July 24th, 2017 and April 19th, 2018. After collecting the snippets, we estimated the sentiment behind the underlying texts by using a service from the IBM Watson family (Ferrucci *et al.*, 2010; Ferrucci *et al.*, 2013) called Natural Language Understanding. Each resulting sentiment score ranges from -1 to 1 (*i.e.*, negative sentiment to positive sentiment). Since many snippets were posted on the same day, we averaged the sentiment scores of all snippets published in a day so as to have a single score per day. In our analysis, we denote the resulting variable by *Scores*. Figure 4 plots a histogram of the obtained sentiment scores. One can immediately see that most snippets associated with Bitcoin are either negative or neutral. For the sake of illustration, consider the snippet *"Unpacking five of the biggest cryptocurrency scams to have hit the crypto world."*, which was posted on April 2018[th], 2018. The resulting sentiment score of -0.74 returned by IBM Watson is very negative due primarily to the role of the word "scams" in that sentence.

Figure 4. Histogram of the Sentiment Scores Associated with Bitcoin Snippets.

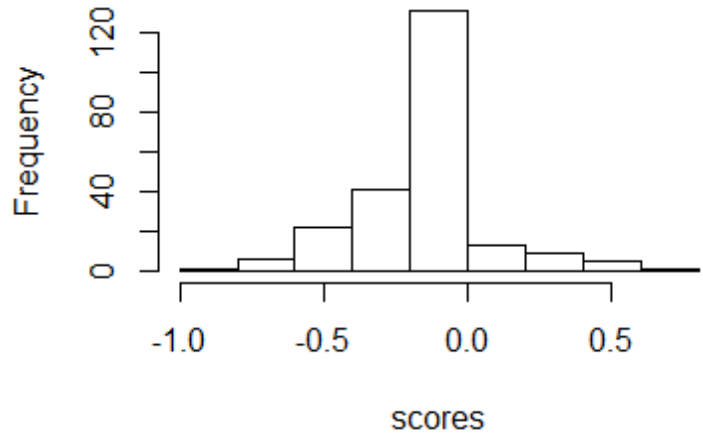

**Final Merged Data**

We note that the first two collected variables, *Bitcoin_Volume* and *Gtrend*, have a total of 270 values, whereas the last two, namely *DJIA_Volume* and *Scores*, have, respectively, 187 and 229. After grouping all variables by day and removing the incomplete cases, we ended up with a data set containing 160 observations and 4 variables. Table 1 illustrates the final data set, which in turn is used in our analysis described next.

**DATA ANALYSIS**

After collecting and preprocessing the data, we next analyze the final data set so as to understand how different variables influence Bitcoin trading activities. In our analysis, we start by reporting some descriptive statistics and correlation matrix in, respectively, Table 2 and 3. From Table 2, one can immediately see that sentiment scores are on average negative, as we already mentioned in the previous section, meaning that most of the collected financial news about Bitcoin are negative in nature. Moreover, the minimum (*6*) and maximum (*85*) *Gtrend* values illustrate that some of the original data points we collected before were lost after merging all the data sets and removing missing data. From the variables *Bitcoin_Volume* and *DJIA_Volume*, one can see that the number of daily Bitcoin transactions is rather small when compared to the number of transactions involving stocks in the DJIA index.

Table 1. Sample of the Final Data.

| Bitcoin_Volume | Gtrend | DJIA_Volume | Scores |
|---|---|---|---|
| 347393 | 29 | 341470000 | -0.761442 |
| 337959 | 33 | 346830000 | 0 |

Table 2. Descriptive Statistics Concerning the Final Data Set.

| Variable | Mean | St. Dev. | Min | Max |
|---|---|---|---|---|
| *Bitcoin_Volume* | 262,529.9 | 69,505.58 | 131,875 | 490,644 |
| *Gtrend* | 18.881 | 14.234 | 6 | 85 |
| *DJIA_Volume* | 370,958,500 | 106,743,091 | 118,610,000 | 823,940,000 |
| *Scores* | -0.117 | 0.235 | -0.761 | 0.665 |

Table 3. Correlation Matrix.

|  | *Bitcoin_Volume* | *Gtrend* | *DJIA_Volume* | *Scores* |
|---|---|---|---|---|
| *Bitcoin_Volume* | -- | 0.614*** | -0.235** | -0.010 |
| *Gtrend* | 0.614*** | -- | 0.214** | 0.025 |
| *DJIA_Volume* | -0.235** | 0.214** | -- | -0.026 |
| *Scores* | -0.010 | 0.025 | -0.026 | -- |
| Note: ** = p-value < 0.01; *** = p-value < 0.001 ||||| 

Focusing now on Table 3, one can see that *Gtrend* and *DJIA_Volume* significantly correlates with *Bitcoin_Volume*. While the former is a positive association, the latter is a negative correlation. This indicates that online popularity, as measured by Google Trends, and Bitcoin trading activity tend to move in the same direction, whereas trading activity associated with the DJIA index and Bitcoin trading activity move in opposite directions. Another interesting fact is that the variable *Scores* is not correlated with any other variable. We return to this point later in the paper. Finally, it is noteworthy that *Gtrend* also positively correlates with *DJIA_Volume*. Recall that the variable *Gtrend* measures the popularity of the term 'Bitcoin' over time. That said, we believe that its positive correlation with *DJIA_Volume* might just be spurious since it seems to contradict the facts that the variables *DJIA_Volume* and *Gtrend* are, respectively, negatively and positively correlated with *Bitcoin_Volume*.

We next extend the above univariate and bivariate analyses by developing a multiple linear regression model where *Bitcoin_Volume* is the dependent variable and all the other variables are independent variables. As one can see from Table 3, the independent variables are not highly correlated, which means that a regression model is unlikely to suffer from multicollinearity issues. Table 4 shows a summary of the obtained regression model. The coefficients in Table 4 confirm what we previously suggested. First, holding everything else constant, Bitcoin trading activities are expected to increase when Bitcoin's online popularity (*Gtrend*) increases. Second, when the number of transactions involving stocks in the DJIA index (*DJIA_Volume*) goes up, the number of Bitcoin transactions are expected to go down. Finally, there is no significant relationship between sentiment scores related to Bitcoin news (*Scores*) and Bitcoin trading activities. The $R^2$ and F-statistic values suggest that our model fits the data well. In particular, it is rather surprising that the three independent variables can explain 51.9% ($R^2$ = 0.519) of the variance in the amount of daily trading activities associated with Bitcoin.

Table 4. Summary of the Multiple Linear Regression Model.

|  | Coefficient | Standard Error | P-value |
|---|---|---|---|
| *(Intercept)* | 289909.95 | 14,245.510 | < 0.001 |
| *Gtrend* | 3403.46 | 277.846 | < 0.001 |
| *DJIA_Volume* | -0.00025 | 0.000037 | < 0.001 |
| *Scores* | -11132.59 | 16,445.560 | 0.499 |
| $R^2$ = 0.519 ||||
| F-statistic = 55.997 (df = 3, 156; p-value < 0.001) ||||

Figure 5. Validating the Assumptions behind the Linear Regression Model. (LEFT) Distribution of the Residuals. (RIGHT) Validating Homoscedasticity.

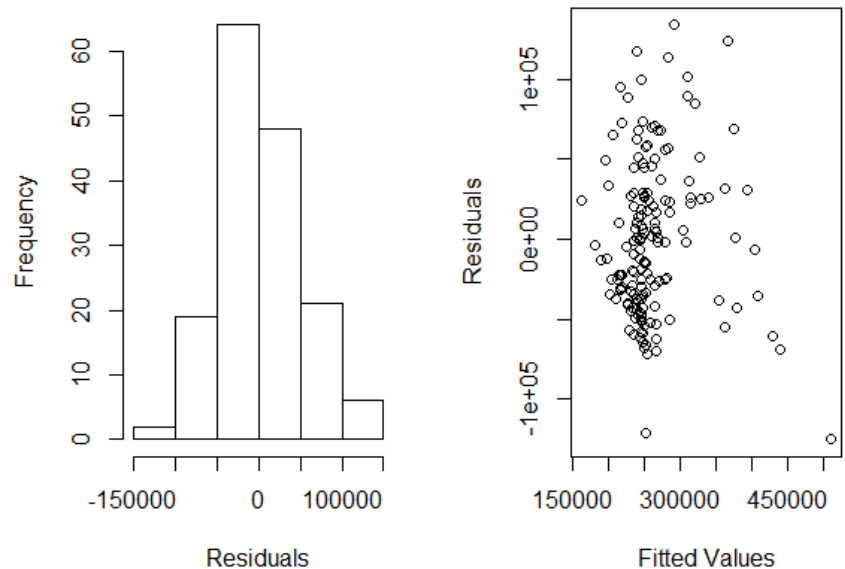

It is important to highlight that we carefully validated the assumptions behind the linear regression model. First, there is strong evidence that the mean of the residuals is equal to zero (one sample t-test; null hypothesis: µ = 0; p-value > 0.999). Second, the distribution of the residuals resembles a normal distribution (see the left part of Figure 5). Finally, the assumption of homoscedasticity seems to hold true (see the right part of Figure 5). Although some observations are flagged as outliers according to traditional guidelines based on Cook's distance, we nonetheless obtained qualitatively the same results when removing those outliers.

**DISCUSSION**

Cryptocurrencies promise to disrupt many traditional industries and the way humans perceive and handle (virtual) money. Given the abundance of cryptocurrencies currently available to the public, it is just natural that only a limited number of virtual coins will eventually prevail. In this paper, we studied some of the factors that might make some coins more popular than others. Specifically, we took the number of daily transactions as a proxy for popularity. This allows one to understand and potentially predict which coins will stand the test of time. Our initial study was focused on Bitcoin since the same is currently the most popular and well-established cryptocurrency. We then investigated how online popularity, trading volume in an unrelated financial market, and financial news influence Bitcoin trading activity.

Our first hypothesis was that online popularity positively correlates with Bitcoin trading volume. Using Google Trends as a proxy for online popularity, we confirm that our first hypothesis is true. Although it was found before that Google Trends values can partially explain Bitcoin prices (Kristoufek, 2013; Kristoufek, 2015), to the best of our knowledge this is the first paper to establish that online popularity also drives Bitcoin trading volume.

Our second hypothesis was that trading volume in non-cryptocurrency financial markets negatively correlates with Bitcoin trading volume. We used the trading volume regarding stocks in the Dow Jones Industrial Average index as a proxy when testing that hypothesis. Our results confirm that the second hypothesis is true in that DJIA trading volume negatively correlates with Bitcoin trading volume. To the best of our knowledge, this is the first work to establish such a relationship between non-cryptocurrency financial markets and cryptocurrency trading volume.

Our last hypothesis was that financial news do not significantly affect Bitcoin trading volume. After collecting data from the most popular Bitcoin public page on Facebook and estimating the sentiment behind the underlying posts using IBM Watson, we confirm that the third hypothesis is also true, *i.e.*, there is no significant relationship between sentiment scores and Bitcoin trading volume. It is fair to acknowledge that the lack of relationship might be due to the fact that, despite having hundreds of thousands of followers, the Facebook page we collected data from might not be influential enough and/or the published news might be somehow biased, *e.g.*, too negative. In hindsight, we recognize that it would be valuable to collect Bitcoin-related news from more than one Facebook page and/or other news sources so as to tackle the abovementioned issues.

We conclude this paper by returning to the discussion in the introductory section, namely how can one know which cryptocurrencies will stand the test of time? Our results suggest very practical guidelines to answer this question. First, one can use Google Trends to track the online popularity of a cryptocurrency over time. When this popularity measure starts going down, then our results imply that trading activities involving the cryptocurrency is also expected to go down and, consequently, the public might be losing interest in the cryptocurrency. Second, one can track the trading volume in different non-cryptocurrency financial markets. When these numbers start going down, then it is expected that trading activities involving cryptocurrencies will go up, meaning that the public might be more interested in cryptocurrencies.

Clearly, the above guidelines rely on the assumption that the results we obtained in this paper are valid for all cryptocurrencies, which is unwise to claim without extra data analyses. That said, besides replicating this study for cryptocurrencies other than Bitcoin, we believe it would be of great value to study how generalizable our results are. For example, would one obtain qualitatively the same results when using stock market indexes other than DJIA or different sources of cryptocurrency-related news? Is there any other way of measuring the online popularity of different cryptocurrencies that perhaps complements Google Trends? We argue that answers to the above questions might be of great value to investors considering to trade cryptocurrencies.